\documentclass[prc,preprint,showpacs,superscriptaddress,floatfix,amsmath,amssymb,nofootinbib]{revtex4}
\usepackage{graphicx}
\usepackage{color}
\usepackage{bm}
 \usepackage[colorlinks,linkcolor=blue,anchorcolor=blue,citecolor=blue]{hyperref}

\newcommand{\br}{{\mathbf{r}}}
\newcommand{\bJ}{{\mathbf{J}}}

\newcommand{\beq}{\begin{equation}}
\newcommand{\eeq}{\end{equation}}
\newcommand{\beqn}{\begin{eqnarray}}
\newcommand{\eeqn}{\end{eqnarray}}

\begin{document}
\title{Impurity effect of Lambda hyperon on collective excitations of atomic nuclei}
\author{J. M. Yao}
\address{School of Physical Science and Technology, Southwest
University, Chongqing 400715, China}
\author{Z. P. Li}
\address{School of Physical Science and Technology, Southwest
University, Chongqing 400715, China}
\author{K. Hagino}
\affiliation{Department of Physics, Tohoku University, Sendai
980-8578, Japan}
\author{M. Thi Win}
\affiliation{Department of Physics, Tohoku University, Sendai
980-8578, Japan}
\author{Y. Zhang}
\affiliation{School of Physics and State Key Laboratory of Nuclear
Physics and Technology, Peking University, 100871 Beijing, China }
\author{J. Meng}
\affiliation{School of Physics and State Key Laboratory of Nuclear
Physics and Technology, Peking University, 100871 Beijing, China }
 \affiliation{School of Physics and Nuclear
Energy, Beihang University, Beijing 100083, China}
\affiliation{Center of Theoretical Nuclear Physics, National
Laboratory of Heavy Ion Accelerator, Lanzhou 730000, China}

\date{\today}

\begin{abstract}

Taking the ground state rotational band in $^{24}$Mg as an example,
we investigate the impurity effect of $\Lambda$ hyperon on
collective excitations of atomic nuclei in the framework of
non-relativistic energy density functional theory. To this end, we
take into account correlations related to the restoration of broken
symmetries and fluctuations of collective variables by solving the
eigenvalue problem of a five-dimensional collective Hamiltonian for
quadrupole vibrational and rotational degrees of freedom. The
parameters of the collective Hamiltonian are determined with
constrained mean-field calculations for triaxial shapes using the
SGII Skyrme force. We compare the low-spin spectrum for $^{24}$Mg
with the spectrum for the same nucleus inside $^{25}_{~\Lambda}$Mg.
It is found that the $\Lambda$ hyperon stretches the ground state
band and reduces the $B(E2:2^+_1 \rightarrow 0^+_1)$ value by $\sim
9\%$, mainly by softening the potential energy surface towards the
spherical shape, even though the shrinkage effect on the average
proton radius is only $\sim0.5\%$.

\end{abstract}


\maketitle

 \section{Introduction}

Since the first discovery of $\Lambda$-hypernuclei by observing
cosmic-rays in emulsion chambers~\cite{Danysz53}, hypernuclei, which
are nuclei with one or more of the nucleons replaced with hyperons,
have been used as a natural laboratory to study hyperon-nucleon and
hyperon-hyperon interactions, properties of hadrons in nuclear
environment, and in particular the impurity effect of hyperon in
nuclear medium~\cite{Chrien89,Dover89,Bando90}. Due to the absence
of Pauli's principle between the nucleon and the $\Lambda$ particle,
a $\Lambda$ hyperon can probe deeply into the interior of nuclear
medium and have important influences on its properties, including
softening the equation of state~\cite{Glendenning00}, modifying the
shape and size of finite nucleus~\cite{Tanida01}, changing the
nuclear binding and thus the driplines of neutrons and
protons~\cite{Samanta06} as well as the fission barrier heights in
heavy nuclei~\cite{Minato09}.

In the past decade, many high-resolution $\gamma$-ray spectroscopy
experiments using germanium detector arrays (Hyperball) have been
carried out for $\Lambda$-hypernuclei~\cite{Hashimoto06} to
understand the nature of $\Lambda$-nucleon interaction in nuclear
medium and the impurity effect of a $\Lambda$ on nuclear structure.
In particular, the facilities built at J-PARC will provide an
opportunity to perform hypernuclear $\gamma$-ray spectroscopy study
with high precision by improving the quality of the secondary
mesonic beam~\cite{Tamura09}. These facilities offer useful tools to
study the low-lying states of hypernuclei, especially those of
medium and heavy hypernuclei. To date, there are many experimental
data not only on the single-$\Lambda$ binding energy but also on the
hypernuclear $\gamma$-ray spectroscopy that allow us to study the
$\Lambda$-nucleon interaction, nuclear medium effects of baryons and
impurity effects induced by a $\Lambda$ hyperon in much greater
detail~\cite{Hashimoto06}.

The theoretical studies for the hypernuclear $\gamma$-ray
spectroscopy are mainly performed with the cluster
model~\cite{Motoba83,Bando90}, few-body
model~\cite{Hiyama03,Nemura02}, and shell model~\cite{Dalitz78}. The
energy level scheme, M1 and E2 transition rates in low-lying states
of light $\Lambda$-hypernuclei have been investigated with either a
one-boson exchange potential or a parameterized spin-dependent
$\Lambda$-nucleon interaction. Due to the numerical difficulty, the
application of these models to medium and heavy hypernuclei is
greatly limited. It is noted that, recently, the framework of
few-body model has been extended to the case of five-body and used
to study the energy levels of the double $\Lambda$-hypernucleus,
$^{11}_{\Lambda\Lambda}$Be~\cite{Hiyama10}.

The framework of nuclear energy-density functionals (EDF) is
nowadays one of the most important microscopic approaches for
large-scale nuclear structure calculations in medium and heavy
nuclei \cite{Bender03} and has already been extended to study
hypernuclei~\cite{Rayet76,Rufa90,Mares94,Schaffner94,Sugahara94,Lv03,Shen06,Vretenar98}.
Recently, both the non-relativistic Skyrme-Hartree-Fock (SHF)
theory~\cite{Zhou07,Schulze10,Win11} and the relativistic mean-field
(RMF) theory~\cite{Win08} have been applied to study the impurity
effect of $\Lambda$ hyperon on the deformation of
$\Lambda$-hypernuclei. The predicted energy surface is somewhat
soft, in which case a large shape fluctuation effect of collective
vibration might be expected. Furthermore, the static
single-reference (SR) EDF is characterized by symmetry breaking
(e.g., translational, rotational, particle number), and can provide
only an approximate description of bulk ground-state properties.
Therefore, to calculate excitation spectra and electromagnetic
transition rates in individual hypernuclei, it is necessary to
extend the SR EDF framework to include collective correlations
related to restoration of broken symmetries and to fluctuations of
collective coordinates.

In recent years several accurate and efficient methods and
algorithms have been developed that perform the restoration of
rotational symmetries in 3D Euler space broken by the static nuclear
mean field and take into account fluctuations around the mean-field
minimum~\cite{Bender08,Yao09,Rodriguez10,Niksic11}. The most
effective approach to configuration mixing calculations is the
generator coordinate method (GCM). Within these methods, the energy
spectrum and electromagnetic transition rates of low-lying excited
states in both light and heavy nuclei have been successfully
reproduced. However, these approaches are currently developed only
for even-even nuclei, which cannot be extended straightforwardly to
study the $\gamma$-ray spectra of single-$\Lambda$ hypernuclei by
simply adding hyperon degree of freedom.

At present, the extension of 3D angular momentum projected GCM
(3DAMP+GCM) method to single-$\Lambda$ hypernuclei based on triaxial
symmetry-breaking intrinsic states is still much complicated and its
applications to medium-heavy and heavy nuclei would be
computationally demanding. As an alternative approach to the 5D
quadrupole dynamics that restores rotational symmetry and allows for
fluctuations around the triaxial mean-field minima, a 5D collective
Bohr Hamiltonian (5DCH) has been formulated with
deformation-dependent parameters determined by microscopic
selfconsistent mean-field
calculations~\cite{Libert99,Prochniak04,Niksic09,Li09}. In this
work, we will construct a 5DCH with the parameters derived from the
Skyrme-Hartree-Fock calculations for the nuclear core in a single
$\Lambda$-hypernucleus and calculate the corresponding low-spin
excitation spectra. The impurity effect of $\Lambda$ hyperon on the
collective motion of an atomic nucleus will be examined by studying
the modifications of collective excitation spectrum. In this way, we
will in this paper concentrate on the modification of the core
nucleus due to the addition of a $\Lambda$ particle, leaving the
evaluation of the spectrum of the whole hypernucleus as a future
work.

The paper is organized as follows. In Section~\ref{sec2} we present
a brief outline of the 5DCH method and the Skyrme-Hartree-Fock
approach for $\Lambda$-hypernucleus. The collective potential energy
surface, parameters in collective Hamiltonian as well as the
resultant collective excitation spectra for $^{24}$Mg and the same
nucleus inside $^{25}_{~\Lambda}$Mg are given in Section~\ref{sec3}.
A brief summary and an outlook for future studies are included in
Section~\ref{sec4}.

 \section{The Method}
 \label{sec2}

 \subsection{Collective Hamiltonian in five dimension}
The collective Hamiltonian that describes the nuclear excitations of
quadrupole vibrations, 3D rotations, and their couplings can be
written in the form:
\begin{equation}
\label{hamiltonian-quant} \hat{H} =
\hat{T}_{\textnormal{vib}}+\hat{T}_{\textnormal{rot}}
              +V_{\textnormal{coll}} \; ,
\end{equation}
where $V_{\textnormal{coll}}$ is the collective potential. The
vibrational kinetic energy reads,
\begin{eqnarray}
\hat{T}_{\textnormal{vib}}
 &=&-\frac{\hbar^2}{2\sqrt{wr}}
   \left\{\frac{1}{\beta^4}
   \left[\frac{\partial}{\partial\beta}\sqrt{\frac{r}{w}}\beta^4
   B_{\gamma\gamma} \frac{\partial}{\partial\beta}\right.\right.\nonumber\\
  && \left.\left.- \frac{\partial}{\partial\beta}\sqrt{\frac{r}{w}}\beta^3
   B_{\beta\gamma}\frac{\partial}{\partial\gamma}
   \right]+\frac{1}{\beta\sin{3\gamma}} \left[
   -\frac{\partial}{\partial\gamma} \right.\right.\nonumber\\
  && \left.\left.\sqrt{\frac{r}{w}}\sin{3\gamma}
      B_{\beta \gamma}\frac{\partial}{\partial\beta}
    +\frac{1}{\beta}\frac{\partial}{\partial\gamma} \sqrt{\frac{r}{w}}\sin{3\gamma}
      B_{\beta \beta}\frac{\partial}{\partial\gamma}
   \right]\right\},
 \end{eqnarray}
and the rotational kinetic energy,
\begin{equation}
\hat{T}_{\textnormal{\textnormal{\textnormal{rot}}}} =
\frac{1}{2}\sum_{\kappa=1}^3{\frac{\hat{J}^2_\kappa}{\mathcal{I}_\kappa}},
\end{equation}
with $\hat{J}_\kappa$ denoting the components of the angular
momentum in the body-fixed frame of a nucleus. It is noted that the
mass parameters $B_{\beta\beta}$, $B_{\beta\gamma}$,
$B_{\gamma\gamma}$, as well as the moments of inertia
$\mathcal{I}_\kappa$, depend on the quadrupole deformation variables
$\beta$ and $\gamma$,
\begin{equation}
 \label{MOI}
\mathcal{I}_\kappa = 4B_\kappa\beta^2\sin^2(\gamma-2\kappa\pi/3),
~~\kappa=1,2,3 \;.
\end{equation}
Two additional quantities that appear in the expression for the
vibrational energy, that is, $r=B_1B_2B_3$, and
$w=B_{\beta\beta}B_{\gamma\gamma}-B_{\beta\gamma}^2 $, determine the
volume element in the collective space. The corresponding eigenvalue
problem is solved by expansion of eigenfunctions in terms of a
complete set of basis functions that depend on the deformation
variables $\beta$ and $\gamma$, and the Euler angles $\phi$,
$\theta$ and $\psi$~\cite{Pro.99}.

The dynamics of the collective Hamiltonian is governed by seven
collective quantities, that is, the collective potential $V_{\rm
coll}$, three mass parameters $B_{\beta\beta}$, $B_{\beta\gamma}$,
and $B_{\gamma\gamma}$, and three moments of inertia
$\mathcal{I}_\kappa$. These quantities are functions of the
intrinsic deformations $\beta$ and $\gamma$ and will be determined
by Skyrme-Hartree-Fock calculations with constraints on the mass
quadrupole moments.

 \subsection{Skyrme-Hartree-Fock approach for $\Lambda$-hypernucleus}

In Ref.~\cite{Win11}, the computer code {\tt ev8}~\cite{Bonche05} of
SHF+BCS approach has already been extended for the study of
$\Lambda$ hypernuclei. Therefore, in the following, we start from
this approach to calculate the seven collective quantities in the
5DCH, as shown in Eq.~(\ref{hamiltonian-quant}).

In the SHF+BCS approach for $\Lambda$ hypernucleus, the total energy
$E$ can be written as the integration of three terms,
 \begin{equation}
 \label{HFE}
 E
 =\int d^3r [{\cal E}_N(\br) + {\cal T}_\Lambda(\br) + {\cal E}_{N\Lambda}(\br)],
 \end{equation}
where ${\cal E}_N(\br)$ is the standard nuclear part of energy
functional, including both $ph$-channel of the Skyrme force and
$pp$-channel of the $\delta$-force, as well as the kinetic energy
density for the nucleons~\cite{Vautherin72,Bonche05}. ${\cal
T}_\Lambda(\br)=\dfrac{\hbar^2}{2m_\Lambda}\tau_\Lambda$ is the
kinetic energy density of $\Lambda$ hyperon. ${\cal
E}_{N\Lambda}(\br)$ is the interaction energy density between the
$\Lambda$ and nucleons given in terms of the $\Lambda$ and nucleon
densities~\cite{Rayet81},
\begin{eqnarray}
  {\cal E}_{N\Lambda}
  &=&
  t^\Lambda_0(1+\dfrac{1}{2}x^\Lambda_0)\rho_\Lambda\rho_N
  +\dfrac{1}{4}(t^\Lambda_1+t^\Lambda_2)(\tau_\Lambda\rho_N+\tau_N\rho_\Lambda)\nonumber\\
  &&+\dfrac{1}{8}(3t^\Lambda_1-t^\Lambda_2)(\nabla\rho_N\cdot\nabla\rho_\Lambda)
  +\dfrac{1}{4}t^\Lambda_3\rho_\Lambda(\rho^2_N+2\rho_n\rho_p)\nonumber\\
  &&+\dfrac{1}{2}W^\Lambda_0(\nabla\rho_N\cdot \bJ_\Lambda+\nabla\rho_\Lambda\cdot \bJ_N)
  \tau_N\rho_\Lambda.
\end{eqnarray}
Here, $\rho_\Lambda, \tau_\Lambda$ and $\bJ_\Lambda$ are
respectively the particle density, the kinetic energy density, and
the spin density of the $\Lambda$ hyperon. These quantities are
given in terms of the single-particle wave-function of $\Lambda$ and
occupation probabilities~\cite{Vautherin72}. $t^\Lambda_0,
t^\Lambda_1, t^\Lambda_2, t^\Lambda_3$, and $W^\Lambda_0$ are the
Skyrme parameters for the $\Lambda$N interaction.

The pairing correlation between the nucleons is taken into account
in the BCS approximation. The density-dependent $\delta$-force is
adopted in the $pp$ channel,
\begin{equation}
 V(\br_1, \br_2)=-g\dfrac{1-\hat P^\sigma}{2}
 \left[1-\dfrac{\rho(\br_1)}{\rho_0}\right]
 \delta(\br_1-\br_2),
\end{equation}
where $\hat P^\sigma$ is the spin-exchange operator, and
$\rho_0=0.16$ fm$^{-3}$.

The HF equations for the nucleons and $\Lambda$ are obtained by
varying the HF energy (\ref{HFE}) with respect to the corresponding
single-particle wave functions and are solved by discretizing
individual single-particle wave functions on a three-dimensional
Cartesian mesh. More details can be found in Ref.~\cite{Bonche05}.

The method of quadratic constraints on the quadrupole moments of the
nuclear density is used to find nuclear intrinsic wave functions
(including the quasiparticle energies $E_i$, occupation
probabilities $v_i$, and single-nucleon wave functions $\psi_i$)
corresponding to the desired quadrupole
deformations~\cite{Bonche05,RS.80}. With these wave functions, one
can calculate the moments of inertia $\mathcal{I}_\kappa$ in Eq.
(\ref{MOI}) using the Inglis-Belyaev formula~\cite{Ing.56,Bel.61}
\begin{equation}
\label{Inglis-Belyaev}
 \mathcal{I}_\kappa =
 \sum_{i,j}{\frac{\left(u_iv_j-v_iu_j \right)^2}{E_i+E_j}
  \langle i |\hat{J}_\kappa | j  \rangle |^2},
\end{equation}
where $\kappa=1, 2, 3$ denotes the axis of rotation, and the
summation of $i, j$ runs over the proton and neutron quasiparticle
states.

The mass parameters $B_{\mu\nu}(\beta,\gamma)$ are also calculated
in the cranking approximation~\cite{GG.79}
\begin{equation}
\label{masspar-B} B_{\mu\nu}(\beta,\gamma)=\frac{\hbar^2}{2}
 \left[\mathcal{M}_{(1)}^{-1} \mathcal{M}_{(3)} \mathcal{M}_{(1)}^{-1}\right]_{\mu\nu}\;,
\end{equation}
with
\begin{equation}
 \label{MMatrix}
 \mathcal{M}_{(n),\mu\nu}(\beta,\gamma)=\sum_{i,j}
 {\frac{\left\langle i\right|\hat{Q}_{2\mu}\left| j\right\rangle
 \left\langle j\right|\hat{Q}_{2\nu}\left| i\right\rangle}
 {(E_i+E_j)^n}\left(u_i v_j+ v_i u_j \right)^2}.
\end{equation}
 The mass parameters $B_{\mu\nu}$ in Eq.(\ref{masspar-B}) can be converted into the
 forms of $B_{\beta\beta}, B_{\beta\gamma}, B_{\gamma\gamma}$ with
 the following relations~\cite{Prochniak09},
 \begin{subequations}\begin{eqnarray}
 B_{\beta\beta} &=& B_{00}a_{00}\cos^2\gamma + 2B_{02}a_{02}\cos\gamma\sin\gamma +
 B_{22}a_{22}\sin^2\gamma,\\
 B_{\beta\gamma} &=& (B_{22}a_{22}-B_{00}a_{00})\cos\gamma\sin\gamma
 +  B_{02}a_{02}(\cos^2\gamma-\sin^2\gamma),\\
 B_{\gamma\gamma} &=& B_{22}a_{22}\cos^2\gamma - 2B_{02}a_{02}\cos\gamma\sin\gamma +
 B_{00}a_{00}\sin^2\gamma,
 \end{eqnarray}\end{subequations}
 where the coefficients $a_{00}, a_{02}, a_{22}$ are as follows,
 \begin{equation}
 a_{00}=\dfrac{9r^4_0A^{10/3}}{16\pi^2},~~
 a_{02}=a_{00}/\sqrt{2},~~
 a_{22}=a_{00}/2,
 \end{equation}
 with $r_0=1.2$.
 \begin{figure}[t]
  \centering
  \includegraphics[width=8cm]{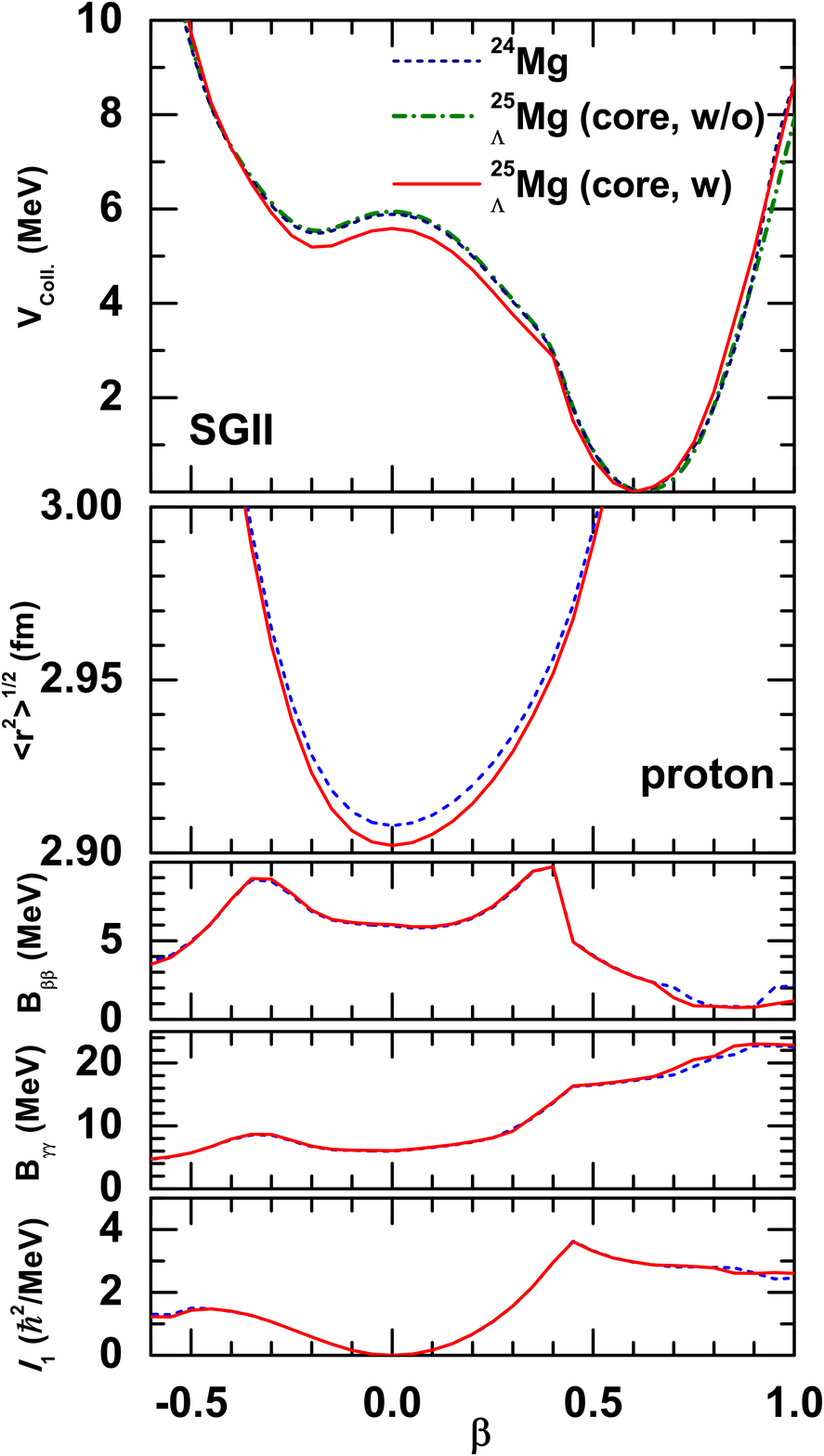}
 \caption{The collective potential $V_{\rm coll}$, the rms radius of protons, the mass parameters
 $B_{\beta\beta}$,  $B_{\gamma\gamma}$,
 and the moment of inertia along the $1$-axis $I_1$ as functions of quadrupole deformation $\beta$ for $^{24}$Mg
 and the nuclear core of $^{25}_\Lambda$Mg from the Skyrme-Hartree-Fock+BCS calculations using the SGII
 force~\cite{Giai81}. The $\Lambda$N interaction energy ${\cal E}_{N\Lambda}$
 has been included ($w$) or excluded ($w/o$) in the collective potential $V_{\rm coll}$ for the
 nuclear core of $^{25}_\Lambda$Mg.}
 \label{fig1}
\end{figure}

The collective potential $V_{\rm coll}$ in the collective
Hamiltonian is obtained by subtracting the zero-point-energy (ZPE)
from the total mean-field energy~\cite{Niksic09},
\begin{equation}
\label{Vcoll} {V}_{\textnormal{coll}}(\beta,\gamma)
 = E_{\textnormal{tot}}(\beta,\gamma)
  - \Delta V_{\textnormal{vib}}(\beta,\gamma) - \Delta
  V_{\textnormal{rot}}(\beta,\gamma),
\end{equation}
where $E_{\textnormal{tot}}$ is the total energy for the nuclear
core in $\Lambda$ hypernucleus. We will investigate two options,
that is, those with ($w$) or without ($w/o$) the interaction part of
energy ${\cal E}_{N\Lambda}$ between the $\Lambda$ and nucleons,
 \begin{eqnarray}
   E_{\textnormal{tot}}
 =\left\{
 \begin{array}{cc}
 \int d^3r {\cal E}_N(\br), &  w/o  \\
  \int d^3r [{\cal E}_N(\br)+ {\cal E}_{N\Lambda}(\br)], &  w \\
 \end{array}
 \right.
 \end{eqnarray}
 In the collective potential $V_{\rm coll}$ of Eq.(\ref{Vcoll}), the vibrational ZPE, $\Delta V_{\textnormal{vib}}$ is given by,
 \begin{equation}
 \Delta V_{\textnormal{vib}}(\beta,\gamma) =
 \dfrac{1}{4}{\rm Tr}[{\cal M}^{-1}_{(3)}{\cal M}_{(2)}],
\end{equation}
where ${\cal M}_{(n),\mu\nu}(\beta,\gamma)$ is determined by
Eq.(\ref{MMatrix}) with the mass quadrupole operators $(\mu,\nu=0,
2)$ defined as ,
\begin{equation}
\hat{Q}_{20}=2z^2-x^2-y^2 \quad \textnormal{and}\quad
\hat{Q}_{22}=x^2-y^2 \;.
\end{equation}
 The rotational part of ZPE is a summation of three terms,
 \begin{equation}
 \Delta V_{\textnormal{rot}}(\beta,\gamma)
 = \sum_{\mu=-2,-1,1} \Delta V_{\mu\mu}(\beta,\gamma),
\end{equation}
with
 \begin{equation}
 \Delta V_{\mu\nu}(\beta,\gamma)
 = \dfrac{1}{4} \dfrac{{\cal M}_{(2),\mu\nu}(\beta,\gamma)}{{\cal
 M}_{(3),\mu\nu}(\beta,\gamma)}.
 \end{equation}
where ${\cal M}_{(n),\mu\nu}(\beta,\gamma)$ is determined by
Eq.(\ref{MMatrix}) with the intrinsic components of quadrupole
operator defined as,
 \begin{eqnarray}
  \hat Q_{2\mu}
 =\left\{
 \begin{array}{cc}
 -2iyz, &  \mu=1  \\
 -2xz, &  \mu=-1 \\
  2ixy, &  \mu=-2 \\
 \end{array}
 \right.
 \end{eqnarray}

 \section{Results and discussion}
 \label{sec3}

Following Ref.~\cite{Win11}, in the $ph$-channel, we adopt the SGII
parameterized Skyrme force~\cite{Giai81} for the NN interaction, and
the No.1 set in Ref.~\cite{Yamamoto88} for the $\Lambda$N
interaction. In the $pp$-channel for nucleons, we follow
Ref.~\cite{Terasaki96} to use $g=1000$ MeV fm$^3$ for both protons
and neutrons. A smooth pairing energy cutoff of 5 MeV around the
Fermi level is used. In the mean-field calculations, the mass
quadrupole moments are constrained to the mesh-points in
$\beta$-$\gamma$ plane with $\beta=0, 0.05, 0.10, \dots, 1.20$ and
$\gamma=0^\circ, 6^\circ, 12^\circ, \dots, 60^\circ$. The $\Lambda$
particle occupies the lowest single-particle state throughout the
constraint calculations. In the following, we take $^{24}$Mg as an
example, and study the impurity effect of $\Lambda$ hyperon by
examining the changes of collective parameters and the resultant
collective excitation spectrum and related obsevables.

 Figure~\ref{fig1} displays the collective potential $V_{\rm coll}$, the rms radius of protons, the mass parameters
 $B_{\beta\beta}$,  $B_{\gamma\gamma}$,
 and the moment of inertia along the $1$-axis $I_1$ as functions of quadrupole deformation $\beta$ for $^{24}$Mg
 and the nuclear core of $^{25}_\Lambda$Mg. The $\Lambda$N interaction energy ${\cal
 E}_{N\Lambda}$ in Eq.(\ref{HFE}) has been included ($w$) or excluded ($w/o$)
 in the collective potential $V_{\rm coll}$ for the
 nuclear core of $^{25}_\Lambda$Mg [cf. Eq.(\ref{Vcoll})]. It is found that the $\Lambda$
 hyperon has negligible influences on the moments of inertia and mass
 parameters of the nuclear core. However, it can lower down the
 barrier in the neighborhood of spherical shape and make the energy
 curve stiffer at large deformed region. In other words, the $\Lambda$
 will reduce the collectivity of $^{24}$Mg, where the $\Lambda$N interaction energy
 plays a major role.

  \begin{figure}[t]
  \centering
  \includegraphics[width=14cm]{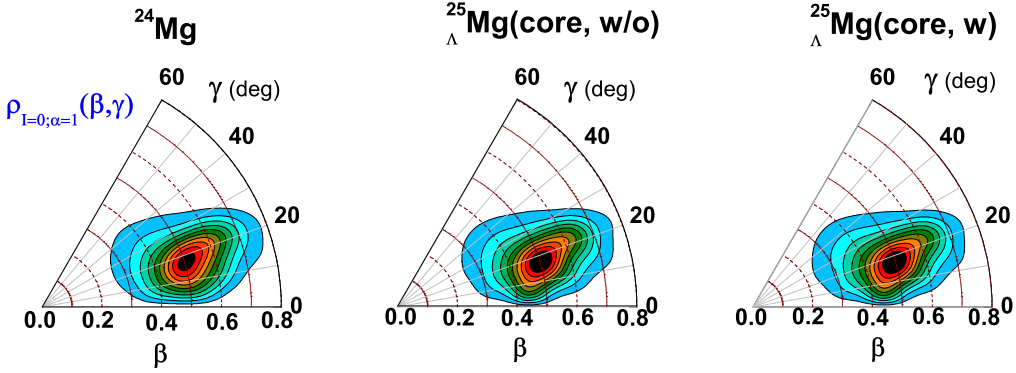}
 \caption{The probability distribution $\rho_{I\alpha}(\beta,\gamma)$ in $\beta$-$\gamma$ plane for the $0^+_1$
 state in $^{24}$Mg (left panel), $^{25}_\Lambda$Mg (core, $w/o$) (middle panel) and $^{25}_\Lambda$Mg (core, $w$) (right panel).}
 \label{fig2}
\end{figure}

 \begin{figure}[t]
  \centering
  \includegraphics[width=10cm]{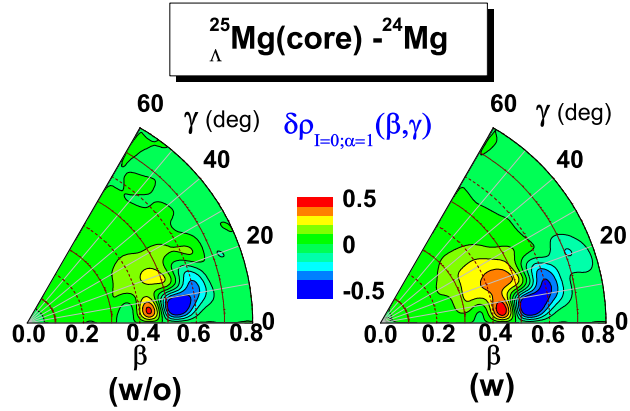}
 \caption{The difference in the probability distribution $\rho_{I\alpha}(\beta,\gamma)$ of $0^+_1$ states between
 $^{25}_\Lambda$Mg (core, $w/o$) and $^{24}$Mg (left panel) as well as
 between $^{25}_\Lambda$Mg (core, $w$) and $^{24}$Mg (right panel).}
 \label{fig3}
\end{figure}

 In Fig.~\ref{fig2}, we plot the probability distribution
 $\rho_{I\alpha}$ in $\beta$-$\gamma$ plane for the $0^+_1$
 state in $^{24}$Mg and the nuclear core of $^{25}_\Lambda$Mg, where the $\rho_{I\alpha}$ is defined as~\cite{Li10},
 \begin{equation}
  \rho_{I\alpha} (\beta,\gamma)
  = \sum_{K}\vert \Psi^{I}_{\alpha,K}(\beta,\gamma)\vert^2 \beta^3 \vert
  \sin3\gamma\vert,
 \end{equation}
 which follows the normalization condition,
 \begin{equation}
  \int^\infty_0 \beta d\beta \int^{2\pi}_0 d\gamma \rho_{I\alpha}(\beta,\gamma) = 1.
 \end{equation}
 Here, $\Psi^{I}_{\alpha,K}(\beta,\gamma)$ is the collective wave function that corresponds to
 the solution of 5DCH in Eq.(\ref{hamiltonian-quant}) and $\alpha=1, 2,\cdots$, labels collective
 eigenstates for a given angular momentum $I$.
 It is shown in Fig.~\ref{fig2} that the $\Lambda$ shifts slightly the probability distribution
 of the $0^+_1$ state to the smaller deformation region. This effect can be seen more
 clearly from the changes in $\rho_{I\alpha}(\beta,\gamma)$
 for the $0^+_1$ state after the introducing of $\Lambda$ hyperon,
 as shown in Fig.~\ref{fig3}, where the differences in the probability distribution $\rho_{0,1} (\beta,\gamma)$
 for the nuclear core of $^{25}_\Lambda$Mg and $^{24}$Mg are
 plotted.  Quantitatively, the average values of $\beta(\gamma)$ are $0.54 (20.0^\circ)$ for $^{24}$Mg
 and these values become $0.53 (20.7^\circ$) for $^{25}$Mg (core, $w/o$),
 and $0.52 (20.8^\circ$) for $^{25}$Mg (core, $w$).

 \begin{figure}[t]
  \centering
  \includegraphics[width=9cm]{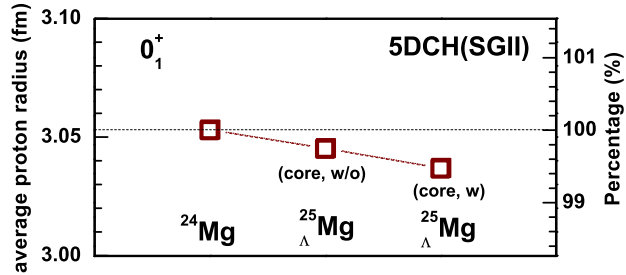}
 \caption{The rms proton radius for $^{24}$Mg and the nuclear core of $^{25}_\Lambda$Mg.}
 \label{fig4}
\end{figure}

 Moreover, it is also shown in Fig.~\ref{fig1} that the rms radius of
 protons is reduced by the $\Lambda$, in particular in the
 neighborhood of spherical shape. However, this shrinkage effect
 on the proton radius of $^{24}$Mg is only $\sim0.5\%$, as illustrated in
 Fig.~\ref{fig4}, where the rms proton radius in $\beta$-$\gamma$ plane for
 both the $^{24}$Mg and the nuclear core of $^{25}_\Lambda$Mg from
 the 5DCH calculations are plotted.

 \begin{figure*}[t]
  \centering
  \includegraphics[width=12cm]{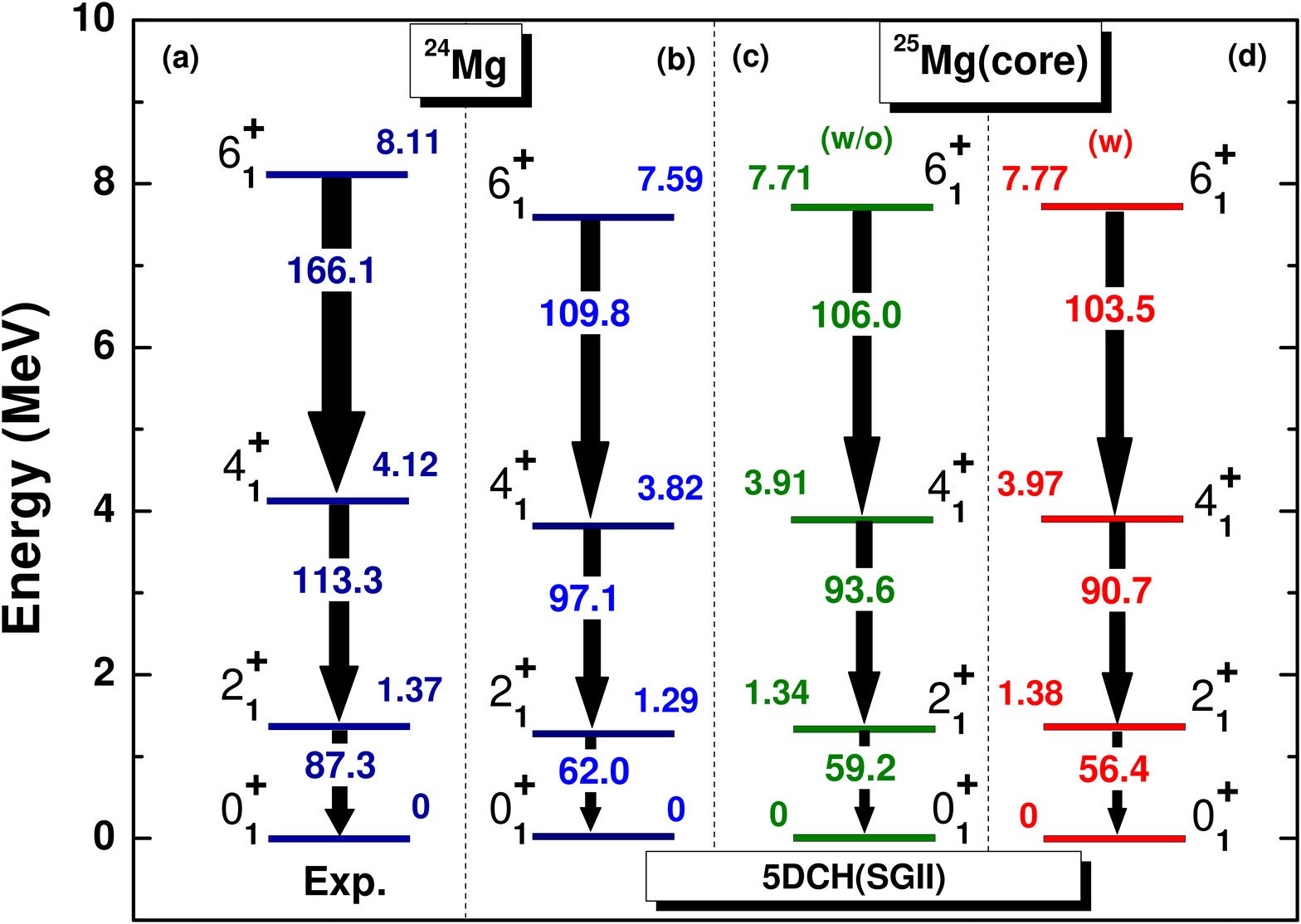}
 \caption{The low-spin spectra of the ground state band for the
 $^{24}$Mg (b) and the nuclear core of $^{25}_\Lambda$Mg (c, d) obtained by the five-dimensional
 collective Hamiltonian (5DCH) with the parameters determined by the
 Skyrme-Hartree-Fock+BCS calculations using the SGII
 force~\cite{Giai81}. The $B(E2)$ values are in units of e$^2$ fm$^4$.
 The spectrum of $^{24}$Mg is compared with the corresponding experimental
 data (a), taken from Ref.~\cite{Endt90}.}
 \label{fig5}
\end{figure*}

 Figure~\ref{fig5} displays the low-spin spectra of ground state band for the
 $^{24}$Mg and the nuclear core of $^{25}_\Lambda$Mg. It is noted that
 the $\Lambda$ stretches the spectra of ground state band. Comparing
 with columns (b) and (d), one finds that the $\Lambda$
 increases the excitation energy of $2^+_1$ state by $\sim 7\%$.
 Moreover, it reduces the E2 transition strength $B(E2: 2^+_1 \rightarrow 0^+_1)$
 by $\sim 9\%$, which is a little smaller than the values,
 $19(4)\%$ or $16(6)\%$ in $^{6}$Li~\cite{Tanida01}.

 \section{Summary and outlook}
 \label{sec4}

The impurity effect of $\Lambda$ hyperon in $^{24}$Mg has been
quantitatively studied in the framework of non-relativistic energy
density functional theory that has been extended to include
correlations related to the restoration of rotational symmetries and
fluctuations of collective variables by solving the eigenvalue
problem of a 5DCH for quadrupole vibrational and rotational degrees
of freedom, with parameters determined by constrained
self-consistent nonrelativistic mean-field calculations for triaxial
shapes using the SGII Skyrme force.  The low-spin spectra for
$^{24}$Mg in both free space and with the additional $\Lambda$ have
been calculated. It has been found that the $\Lambda$ hyperon shifts
the collective wave function of ground state to a smaller
deformation region by softening the nuclear collective potential
surface in the neighborhood of spherical shape. As the consequence
of this effect, the spectra of ground state band becomes stretched
and the excitation energy of $2^+_1$ state is increased by $\sim
7\%$. Moreover, the $B(E2: 2^+_1 \rightarrow 0^+_1)$ value is
reduced by $\sim 9\%$.  However, the shrinkage effect on the average
proton radius is found to be only $\sim0.5\%$.

As pointed out in Refs.~\cite{Win08,Schulze10,Win11}, the influence
of the addition of $\Lambda$ particle might be stronger in the
relativistic mean-field approach. Therefore, it would be very
interesting to extend this work to the relativistic case. In
addition, to calculate directly the $\gamma$-spectra of
single-$\Lambda$ hypernucleus, one has to extend the current EDF
based 3DAMP+GCM or 5DCH models for the odd-mass or odd-odd nucleus.
Working along this direction is in progress.

 \begin{acknowledgments}
We would like to thank P. Z. Ning and T. Koike for helpful
discussions. K.H. acknowledges the Global Center of Excellence
Program "Weaving Science Web beyond Particle-Matter Hierarchy" at
Tohoku University for financial support and thanks the Southwest
University for its hospitality. This work is partly supported by the
Major State 973 Program 2007CB815000 and the NSFC under Grants No.
10947013 and No. 10975008; the Fundamental Research Funds for the
Central Universities (XDJK2010B007); the Southwest University
Initial Research Foundation Grant to Doctor (No. SWU109011 and No.
SWU110039); and the Japanese Ministry of Education, Culture, Sports,
Science and Technology by Grant-in-Aid for Scientific Research under
Program No. 22540262.
 \end{acknowledgments}

 \begin{appendix}
 \section{Calculations of moments of inertia with the EV8 code}
 \label{Appendix}
 In the {\tt ev8} code~\cite{Bonche05}, the single-particle (s.p.) wave-function of $k$-state $\Phi_{k}(\br)$,
 discretized on a three-dimensional Cartesian mesh, is written in the $4$-component
 form,
  \beqn
  \Phi_{k}=\begin{pmatrix}
  \Psi^{(1)}_k+i \Psi^{(2)}_k \\
  \Psi^{(3)}_k+i  \Psi^{(4)}_k
  \end{pmatrix}
 \eeqn
 where $\Psi^{(\alpha)}_k$ ($\alpha=1,2,3,4$) are real functions corresponding to
 the real and imaginary, spin-up and spin-down parts of $\Phi_{k}$.
 The time-reversed state of $\Phi_{k}$ are determined by
  \beqn
  \Phi_{\bar k}
  \equiv\hat T \Phi_{k}
  =
  -\begin{pmatrix}
  \Psi^{(3)}_k -i  \Psi^{(4)}_k \\
  -\Psi^{(1)}_k+i  \Psi^{(2)}_k \\
  \end{pmatrix}.
 \eeqn
 Therefore, the components in $\Phi_{\bar k}$
 are connected with the components in $\Phi_{k}$
 by the following relations,
 \beq
 \Psi^{(1)}_{\bar k} =-\Psi^{(3)}_k,~
 \Psi^{(2)}_{\bar k} = \Psi^{(4)}_k,~
 \Psi^{(3)}_{\bar k} = \Psi^{(1)}_k,~
 \Psi^{(4)}_{\bar k} =-\Psi^{(2)}_k.
 \eeq

 \begin{table}
 \tabcolsep=20pt
 \caption{Parities of four components $\Psi^{(\alpha)}_k$ ($\alpha=1,2,3,4$)
 in single-particle wave function $\Phi_{k}$ of $k$-state with respect to
 the planes $x=0, y=0, z=0$. The parity of $k$-state is denoted as $p_k$.}
  \begin{tabular}{c|ccc}
    \hline\hline
                &  $x$ & $y$ & $z$ \\
    \hline
 $\Psi^{(1)}_k$ &  $+$ & $+$ & $p_k$ \\
 $\Psi^{(2)}_k$ &  $-$ & $-$ & $p_k$\\
 $\Psi^{(3)}_k$ &  $-$ & $+$ & $-p_k$\\
 $\Psi^{(4)}_k$ &  $+$ & $-$ & $-p_k$\\
    \hline\hline
  \end{tabular}
 \label{Symmetry}
  \end{table}

  Together with the parities of the four components $\Psi^{(\alpha)}_k$ in single-particle wave function
  $\Phi_{k}$ of $k$-state with respect to the planes $x=0, y=0, z=0$, as shown in Table ~\ref{Symmetry},
  the moments of inertia $I_{1,2,3}$ in Eq.(\ref{Inglis-Belyaev})
  can be simplified as
 \begin{subequations}
  \beqn
 I_{1,2}
 &=&2\sum_{i,j>0}\dfrac{(u_iv_j-v_iu_j)^2}{E_i+E_j}
 \left\vert\langle \bar i\vert\hat J_{1,2}\vert j\rangle\right\vert^2,\\
 I_3
 &=&2\sum_{i,j>0}\dfrac{(u_iv_j-v_iu_j)^2}{E_i+E_j}
 \left\vert\langle i\vert\hat J_3\vert j\rangle\right\vert^2,
 \eeqn\end{subequations}
 where the s.p. states $i, j$ have the same parities ($p_i=p_j$) and
 the non-zero matrix elements are determined by,

 \begin{subequations}\beqn
  \langle \bar i\vert\hat J_{1}\vert j\rangle
 &=&\int\int\int^{+\infty}_{-\infty} dxdydz\nonumber\\
 && \left[-\Psi^{(3)}_{i}(\dfrac{1}{2}\Psi^{(3)}_j+i\hat L_x\Psi^{(2)}_j)
   +\Psi^{(4)}_{i}(\dfrac{1}{2}\Psi^{(4)}_j-i\hat  L_x\Psi^{(1)}_j)\right.\nonumber\\
 &&+\left. \Psi^{(1)}_{i}(\dfrac{1}{2}\Psi^{(1)}_j+i\hat  L_x\Psi^{(4)}_j)
   - \Psi^{(2)}_{i}(\dfrac{1}{2}\Psi^{(2)}_j-i\hat
   L_x\Psi^{(3)}_j)\right],\\
 \langle \bar i\vert\hat J_2\vert j\rangle
 &=&\int\int\int^{+\infty}_{-\infty} dxdydz\nonumber\\
 && \left[-\Psi^{(3)}_{i}(-\dfrac{1}{2}\Psi^{(3)}_j-i\hat L_y\Psi^{(1)}_j)
   + \Psi^{(4)}_i(-\dfrac{1}{2}\Psi^{(4)}_j-i\hat  L_y\Psi^{(2)}_j)\right.\nonumber\\
 &&+\left.\Psi^{(1)}_i(\dfrac{1}{2}\Psi^{(1)}_j-i\hat  L_y\Psi^{(3)}_j)
   -\Psi^{(2)}_i(\dfrac{1}{2}\Psi^{(2)}_j-i\hat
   L_y\Psi^{(4)}_j)\right],\\
 \langle i\vert\hat J_3\vert j\rangle
 &=&\int\int\int^{+\infty}_{-\infty} dxdydz\nonumber\\
 &&\left[+\Psi^{(1)}_i(\dfrac{1}{2}\Psi^{(1)}_j  +i\hat  L_z\Psi^{(2)}_j)
 +\Psi^{(2)}_i  (\dfrac{1}{2}\Psi^{(2)}_j  -i\hat  L_z\Psi^{(1)}_j)\right.\nonumber\\
 &&+\left.\Psi^{(3)}_i(-\dfrac{1}{2}\Psi^{(3)}_j +i\hat  L_z\Psi^{(4)}_j)
 +\Psi^{(4)}_i  (-\dfrac{1}{2}\Psi^{(4)}_j -i\hat
 L_z\Psi^{(3)}_j)\right].
 \eeqn\end{subequations}
 In the above equations, $\hat L_\kappa$ ($\kappa=x,y,z$) denote the components of orbital angular momentum
 operator.

 \end{appendix}

\end{document}